
\magnification=1200
\hsize=16.5truecm
\vsize=23truecm
\hfuzz=30pt
\font\title=cmr10 scaled \magstep4
\font\abstract=cmr8
\font\SC=cmcsc10   
\font\dick=msbm10
\font\sect=cmbx10 scaled \magstep2
\font\subsect=cmbx10 scaled \magstep1
\font\typewrite=cmtt10

\def\square{{\vcenter{\vbox{\hrule height.4pt \hbox{\vrule width.4pt
height1.45ex \kern1.45ex \vrule width.4pt}
\hrule height.4pt}}}}

\def\qed{\hfill$\square$\par\medskip\noindent}

\def\today{\ifcase\month\or January \or \February\or March \or April\or May
\or June \or July\or August\or September\or October\or November\or December\fi
~\number\day, \number\year}


\def\proof{\medskip\noindent{\it Proof.} }
\def\proofof #1.{\medskip\noindent{\it Proof of #1.} }

\def\boldhead#1 {\bigbreak\medskip\noindent{\bf #1.}
                \par\medskip\nobreak\noindent}
\def\section#1:#2. {\bigbreak\bigskip\noindent{\sect#1. #2.}
                 \nobreak\medskip\nobreak\noindent}
\def\subsection#1:#2. {\bigbreak\medskip\noindent{\subsect#1. #2.}
                 \nobreak\smallskip\nobreak\noindent}
\def\newsection#1 {\bigbreak\bigskip\noindent\subsecno=0{\sect\secn.\quad#1.}
                 \nobreak\medskip\nobreak\noindent}
\def\newsubsection#1 {\bigbreak\medskip\noindent{\subsect\number\secno.\subsecn.\quad#1.}
                 \nobreak\smallskip\nobreak\noindent}
\def\references{\bigbreak\bigskip\noindent{\bf References}
                 \medskip\nobreak\noindent}
\def\acknowledgement{\bigbreak\bigskip\noindent{\bf Acknowledgement}
                 \medskip\nobreak\noindent}

\def\i #1 {\item{{\rm#1}}}

\def\sn{\smallskip\noindent}
\def\mn{\medskip\noindent}

\def\pn{\noindent}


\def\A{{\cal A}}

\def\C{{\cal C}}
\def\D{{\cal D}}

\def\H{{\cal H}}

\def\AA{{\hbox{\dick A}}}
\def\NN{{\hbox{\dick N}}}

\def\RR{{\hbox{\dick R}}}
\def\CC{{\hbox{\dick C}}}


\def\a{\alpha}
\def\b{\beta}
\def\g{\gamma}

\def\th{\vartheta}

\def\l{\lambda}

\def\ph{\varphi}
\def\oo{\omega}
\def\G{\Gamma}
\def\D{\Delta}


\def\Im{{\rm Im}\,}

\def\and{\quad{\rm and}\quad}

\def\coth{{\rm coth}\,}
\def\aut{{\rm aut}\,}

\def\and{\quad{\rm and}\quad}


\def\to{\rightarrow}
\def\tto{\longrightarrow}

\def\li#1{\lim_{#1\to\infty}}

\def\szi#1{\sum_{#1=0}^\infty}
\def\izi{\int_0^\infty}
\def\ini#1{\int_{#1}^\infty}

\def\pon#1{\prod_{#1=1}^n}
\def\tuple#1_#2{#1_1,#1_2,\cdots,#1_{#2}}

\def\seqz#1{#1_0,#1_1,#1_2,\cdots}

\def\one{{\bf 1}}

\def\Implies{\quad\Longrightarrow\quad}
\def\Equivalent{\quad\Longleftrightarrow\quad}

\def\intR{\int_{-\infty}^\infty}
\def\intRp{\int_0^\infty}

\def\inp#1#2{\langle#1,#2\rangle}

\def\set#1#2{\bigl\{\>#1\>\big|\>#2\>\bigr\}}
\def\norm#1{\|#1\|}

\def\bnorm#1{\bigl\|#1\bigr\|}


\def\mapright#1{\smash{\mathop{\longrightarrow}\limits^{#1}}}
\def\mapdown#1{\Big\downarrow\rlap{$\vcenter{\hbox{$\scriptstyle#1$}}$}}
\def\mapup#1{\llap{$\vcenter{\hbox{$\scriptstyle#1$}}$}\Big\uparrow}
\def\diagram#1#2#3#4#5#6#7#8
{$$\matrix{#3&\mapright{#4}&#5\cr
   \noalign{\vskip4pt}
   \mapup{#2}&&\mapdown{#6}\cr
   \noalign{\vskip4pt}
   #1&\mapright{#8}&#7\cr}$$}

\def\line#1{\hbox to\hsize{\hskip\leftskip#1\hskip\rightskip}}

\def\do{\downarrow}

\def\half{{\textstyle{1\over2}}}

\def\sixth{{\textstyle{1\over6}}}



\def\sCn{\sum_{c_1=0}^0\sum_{c_2=0}^1\sum_{c_3=0}^2\cdots\sum_{c_n=0}^{n-1}}
\def\sCns{\sum_{c_1=0}^0\cdots\sum_{c_n=0}^{n-1}}
\def\iDn#1{\mathrel{\mathop{\int\cdots\int}_{0\le t_1\le\cdots\le t_n}}
           #1dt_1dt_2\cdots dt_n}
\def\iDthree#1{\mathrel{\mathop{\int\int\int}_{0\le t_1\le t_2\le t_3}}
           #1dt_1dt_2dt_3}

\def\at{{\widetilde\alpha}}

\def\gt{{\widetilde\gamma}}
\def\xit{{\widetilde\xi}}

\def\fet{{\widetilde f}}

\def\fest{{\tilde f}}

\def\mt{{\widetilde m}}
\def\wt{{\widetilde w}}

\def\mup{\mu_{+}}

\def\ot{{\widetilde \omega}}

\def\Gt{{\widetilde G}}



\countdef\eqnno=99
\def\eqn{{\global\advance\eqnno by 1}\eqno(\number\eqnno)}
\countdef\lemno=98
\def\lemn{{\global\advance\lemno by 1}\number\lemno}
\countdef\secno=97
\def\secn{{\global\advance\secno by 1}\number\secno}
\countdef\subsecno=96
\def\subsecn{{\global\advance\subsecno by 1}\number\subsecno}

\newread\epsffilein    
\newif\ifepsffileok    
\newif\ifepsfbbfound   
\newif\ifepsfverbose   
\newdimen\epsfxsize    
\newdimen\epsfysize    
\newdimen\epsftsize    
\newdimen\epsfrsize    
\newdimen\epsftmp      
\newdimen\pspoints     
\newdimen\hormaat      
\newdimen\vermaat      
\pspoints=1bp          
\epsfxsize=0pt         
\epsfysize=0pt         
\def\epsfbox#1{\global\def\epsfllx{72}\global\def\epsflly{72}%
   \global\def\epsfurx{540}\global\def\epsfury{720}%
   \def\lbracket{[}\def\testit{#1}\ifx\testit\lbracket
   \let\next=\epsfgetlitbb\else\let\next=\epsfnormal\fi\next{#1}}%
\def\epsfgetlitbb#1#2 #3 #4 #5]#6{\epsfgrab #2 #3 #4 #5 .\\%
   \epsfsetgraph{#6}}%
\def\epsfnormal#1{\epsfgetbb{#1}\epsfsetgraph{#1}}%
\def\epsfgetbb#1{%
%
%
\openin\epsffilein=#1
\ifeof\epsffilein\errmessage{I couldn't open #1, will ignore it}\else
%
%
   {\epsffileoktrue \chardef\other=12
    \def\do##1{\catcode`##1=\other}\dospecials \catcode`\ =10
    \loop
       \read\epsffilein to \epsffileline
       \ifeof\epsffilein\epsffileokfalse\else
%
%
          \expandafter\epsfaux\epsffileline:. \\%
       \fi
   \ifepsffileok\repeat
   \ifepsfbbfound\else
    \ifepsfverbose\message{No bounding box comment in #1; using defaults}\fi\fi
   }\closein\epsffilein\fi}%
%
%
\def\epsfclipstring{}
\def\epsfsetgraph#1{%
   \epsfrsize=\epsfury\pspoints
   \advance\epsfrsize by-\epsflly\pspoints
   \epsftsize=\epsfurx\pspoints
   \advance\epsftsize by-\epsfllx\pspoints
%
%
   \epsfxsize\epsfsize\epsftsize\epsfrsize
   \ifnum\epsfxsize=0 \ifnum\epsfysize=0
      \epsfxsize=\epsftsize \epsfysize=\epsfrsize
      \epsfrsize=0pt
%
%
     \else\epsftmp=\epsftsize \divide\epsftmp\epsfrsize
       \epsfxsize=\epsfysize \multiply\epsfxsize\epsftmp
       \multiply\epsftmp\epsfrsize \advance\epsftsize-\epsftmp
       \epsftmp=\epsfysize
       \loop \advance\epsftsize\epsftsize \divide\epsftmp 2
       \ifnum\epsftmp>0
          \ifnum\epsftsize<\epsfrsize\else
             \advance\epsftsize-\epsfrsize \advance\epsfxsize\epsftmp \fi
       \repeat
       \epsfrsize=0pt
     \fi
   \else \ifnum\epsfysize=0
     \epsftmp=\epsfrsize \divide\epsftmp\epsftsize
     \epsfysize=\epsfxsize \multiply\epsfysize\epsftmp   
     \multiply\epsftmp\epsftsize \advance\epsfrsize-\epsftmp
     \epsftmp=\epsfxsize
     \loop \advance\epsfrsize\epsfrsize \divide\epsftmp 2
     \ifnum\epsftmp>0
        \ifnum\epsfrsize<\epsftsize\else
           \advance\epsfrsize-\epsftsize \advance\epsfysize\epsftmp \fi
     \repeat
     \epsfrsize=0pt
    \else
     \epsfrsize=\epsfysize
    \fi
   \fi
%
%
   \ifepsfverbose\message{#1: width=\the\epsfxsize, height=\the\epsfysize}\fi
   \epsftmp=10\epsfxsize \divide\epsftmp\pspoints
   \vbox to\epsfysize{\vfil\hbox to\epsfxsize{%
      \ifnum\epsfrsize=0\relax
        \includegraphics{#1}%
      \else
        \epsfrsize=10\epsfysize \divide\epsfrsize\pspoints
        \includegraphics{#1}%
      \fi
      \hfil}}%
\global\hormaat=\epsfxsize \global\vermaat=\epsfysize
\global\epsfxsize=0pt\global\epsfysize=0pt}%
%
%
{\catcode`\%=12 \global\let\epsfpercent=
%
%
\long\def\epsfaux#1#2:#3\\{\ifx#1\epsfpercent
   \def\testit{#2}\ifx\testit\epsfbblit
      \epsfgrab #3 . . . \\%
      \epsffileokfalse
      \global\epsfbbfoundtrue
   \fi\else\ifx#1\par\else\epsffileokfalse\fi\fi}%
%
%
\def\epsfempty{}%
\def\epsfgrab #1 #2 #3 #4 #5\\{%
\global\def\epsfllx{#1}\ifx\epsfllx\epsfempty
      \epsfgrab #2 #3 #4 #5 .\\\else
   \global\def\epsflly{#2}%
   \global\def\epsfurx{#3}\global\def\epsfury{#4}\fi}%
%
%
\def\epsfsize#1#2{\epsfxsize}
%
%



\epsfverbosetrue

\def\appict #1#2 {\epsfxsize=#1 \medbreak\centerline{\epsfbox{#2}}\medbreak}

\def\appictname #1#2#3 { \epsfxsize=#1
    \medbreak \centerline{\epsfbox{#2}} \centerline{#3} \medbreak }

\def\indpict #1#2#3 {\medbreak\noindent \hbox to \hsize{\epsfxsize=#1
    \epsfbox{#2} \hfil \vbox to \the\vermaat
    {\advance\hsize by -\hormaat \advance\hsize by -1.2cm \vfil #3 \vfil  }
    } \medbreak }

\def\indpictname #1#2#3#4 {\medbreak\noindent \hbox to \hsize{
    \vbox{\hbox{\epsfxsize=#1  \epsfbox{#2}} \hbox to \hormaat{\hfil #3 \hfil} }
    \hfil \advance\vermaat by \baselineskip \vbox to \vermaat
    {\advance\hsize by -\hormaat \advance\hsize by -1.2cm \vfil #4 \vfil  }
    } \medbreak }

\def\rindpict #1#2#3 {\medbreak\noindent  \hbox to \hsize{\epsfxsize=#1
    \hskip\hsize\hskip-\epsfxsize \epsfbox{#2} \hskip\epsfxsize\hskip-\hsize
    \vbox to \the\vermaat
    {\advance\hsize by -\hormaat \advance\hsize by -1.2cm \vfil #3 \vfil  }  
    }  \medbreak }

\def\rindpictname #1#2#3#4 {\medbreak\noindent \hbox to \hsize{\epsfxsize=#1
    \hskip \hsize \hskip -\epsfxsize
    \vbox{\hbox {\epsfbox{#2}}\hbox to \hormaat {\hfil #3 \hfil}}\hskip -\hsize
    \advance\vermaat by \baselineskip
    \vbox to \the\vermaat
       {\advance\hsize by -\hormaat \advance\hsize by -1.2cm
        \vfil #4 \vfil } 
    } \medbreak  }

\def\inlpict #1#2 {$ \vcenter {\hbox{\epsfxsize=#1 \epsfbox{#2} } } $ \hskip .5em}

\def\lmarpict #1#2 {\epsfxsize=#1 \par \vskip \baselineskip
    \llap{\epsfbox{#2} \hskip 1em}
    \vskip -\vermaat \vskip -\baselineskip }

\def\rmarpict #1#2 {\par \vskip \baselineskip \hskip -\parindent\hskip\hsize
    \rlap{\epsfxsize=#1 \hskip 1em \epsfbox{#2}}
    \vskip -\vermaat \vskip -\baselineskip }


\def\newline{\par\noindent\vskip-\parskip}

\def\lddots{\mathinner{\mkern1mu
\raise1pt\vbox{\kern7pt\hbox{.}}\mkern2mu
\raise4pt\hbox{.}\mkern2mu\raise7pt\hbox{.}\mkern1mu}}
\def\stop{\vbox{\hrule height .5pt
 \hbox{\vrule width .5pt height 5pt \kern 5pt \vrule width .5pt}
 \hrule height .5pt}}

\def\boxalinea#1  {\smallbreak\centerline {\hbox
{\vbox  {\hrule\hbox{\vrule\kern3pt\vbox{\kern3pt
\hbox {\vbox{\hsize .90\hsize #1}}\kern3pt}\kern3pt\vrule}\hrule}  }  }  }

\def\boxalineavet#1  {\smallbreak\centerline {\hbox
{\vbox  {\hrule height 1pt\hbox{\vrule width 1pt\kern3pt\vbox{\kern3pt
\hbox {\vbox{\hsize .90\hsize #1}}\kern3pt}\kern3pt\vrule width 1pt}
\hrule height 1pt}  }  }  }

\def\boxit#1  {\smallbreak\centerline {\hbox
{\vbox  {\hrule\hbox{\vrule\kern3pt\vbox{\kern3pt
\hbox {#1}\kern3pt}\kern3pt\vrule}\hrule}  }  }  }


\def\proofsec{4}
\def\resultsec{2}
\def\examplesec{3}
\def\packing{5}

\nopagenumbers

\vglue 1.5truecm \centerline{\title A Galton-Watson estimate}\bigskip
\centerline{\title for Dyson series}
\bigskip\bigskip\bigskip
\centerline{\SC Hans Maassen}
\bigskip
\settabs\+Some more space in front\qquad&\cr
        \+&Department of Mathematics\cr
        \+&Radboud University Nijmegen\cr
        \+&Heyendaalseweg 135\cr
        \+&6525 AJ Nijmegen\cr
        \+&The Netherlands\cr
\smallskip
        \+&{\typewrite maassen@math.ru.nl}\cr
\bigskip\medskip
\centerline{\SC Dmitri Botvich}
\bigskip
\settabs\+Some more space in front\qquad&\cr
        \+&Telecommunications Software \& Systems Group\cr
        \+&Waterford Institute of Technology\cr
        \+&Carriganore Campus\cr
        \+&co. Waterford\cr
        \+&Ireland\cr
\smallskip
        \+&{\typewrite dbotvich@tssg.org}\cr
\bigskip
\centerline{\today}

\vglue 1.5truecm

{\narrower\noindent{\bf Abstract:}\par\noindent
We consider the question of convergence of particular series
of integrals, which are labeled by rooted trees.
Necessary and sufficient criteria for convergence are obtained,
together with an explicit expression for the sum.
The technique used is strongly reminiscent of the generating
function approach of Galton and Watson to branching processes.
The interest in these series derives from the Dyson series expansion
for the perturbation of a free quantum dynamics by a local potential:
the convergence of the series imlies that the perturbed dynamics exists
and is unitarily equivalent with the free one.
\par}

\bigskip\bigskip\noindent
{\sl We dedicate this paper to John T. Lewis, who was a teacher and source
of inspiration to both of us.}

\vfill\eject

\pageno=1 \footline={\hss--~\folio~--\hss}
\newsection{Introduction} \edef\introsec{\number\secno}\noindent
The purpose of this paper is to provide a technical result that can be
used in the perturbation theory of infinite quantum systems.
It provides a symmability condition, and an upper bound,
for the Dyson series associated to local perturbations of
well-behaved `free' time evolutions.

\noindent
It is of direct relevance in the quasi-particle description of bosonic
and fermionic systems [BoM83], the study of approach to thermal equilibrium
in Caldeira-Leggett type models [Maa84] and anharmonic chrystals
[FiL99, FiL03], and the description of Rayleigh scattering [Spo97]
and dissipative transport [FMU04].

\noindent
The result has been available in preprint form for some time [BGM99],
and has already been applied in the derivation of Green-Kubo
formulas and Onsager
reciprocity relations for coupled Fermi systems [JOP06].
Here we give a formulation and a proof with only a sketch of the context.

\smallskip\noindent
{\bf Asymptotic Completeness}\smallskip\noindent
The key issue in the applications is asymptotic completeness:
if the scattering operator associated to the perturbation is onto,
it establishes an equivalence between the free and the perturbed
dynamics.
The ergodic properties of the free evolution are then
preserved by the perturbation.

\noindent
This scattering approach to infinite quantum systems was initiated by
Robinson [Rob73]; we give a sketch here for the $C^*$ situation;
with some care it extends to the von Neumann algebra context,
where the continuity assumptions are less restrictive.

\noindent
A {\it quantum dynamical system} is a triple $(\A,\oo,\a_t)$,
where $\A$ is a $C^*$-algebra,
$\oo$ is a state on $\A$, and $(\a_t)_{t\in\RR}$
a strongly continuous one-parameter group of *-automorphisms of $\A$.
By the Gel'fand-Naimark-Segal construction the pair $(\A,\oo)$
determines a Hilbert space $\H$, a unit vector $\xi\in\H$,
and a representation of $\A$ as an algebra of bounded operators
on $\H$ such that $\A\xi$ is dense in $\H$ and
$\inp{\xi}{A\xi}=\oo(A)$ for all $A\in\A$.
Let us assume that the dynamical system $(\A,\oo,\a_t)$ is 
{\it mixing}, i.e. for all $A\in\A$ and all unit vectors
$\psi\in\H$ we have:
   $$\li t\inp\psi{\a_t(A)\psi}=\oo(A)\;.\eqn$$
\edef\mixing{\number\eqnno}\noindent
The dynamics $\a_t$ determines a one-parameter group of unitary operators
$(U_t)_{t\in\RR}$ on $\H$ by the relation
$U_t A\xi=\a_t(A)\xi$.
The generator $H$ of this group, given by
   $U_t=e^{itH}$,
is the {\it Hamiltonian} of the quantum dynamical system.
Now let some self-adjoint element $V$ of $\A$ be given,
and let us define new dynamics on $\A$ by
   $$\at_t:\A\to\A: A\mapsto e^{it(H+V)} A e^{-it(H+V)}\;.$$
In order to compare the two evolutions $\a_t$ and $\at_t$,
Robinson proposed to consider the scattering operator $\g$
given by the pointwise norm limit
   $$\g(A):=\li t \at_{-t}\circ\a_t(A)\;,$$
which exists under the fairly mild integrability condition
   $$\int_0^\infty\norm{[\a_t(V),A]}dt<\infty\;.$$
The operator $\g$ intertwines the two evolutions:
   $$\g\circ\a_t(A)=\at_t\circ\g(A)\;,\quad(t\in\RR,A\in\A)\;.$$
Suppose further that there is another vector $\xit\in\H$ such that the
state $\ot:A\mapsto\inp{\xit}{A\xit}$ is invariant for the perturbed
evolution,
and that the space of vectors $A\xi$ for which $\g(A)$ exists,
is dense in $\H$.
Then it is not difficult to show (cf. [Maa82]) that
the operator $\G_0:A\xi\mapsto\g(A)\xit$ extends to an isometry
$\G:\H\to\H$.
In general, $\G$ need not be unitary, but if it is,
the scattering operator is said to be {\it asymptotically complete},
and the two evolutions $\a_t$ and $\at_t$ are unitarily equivalent.

\noindent
For the invertibility of $\G$ it suffices that the inverse limit
   $$\gt(A):=\li t\a_{-t}\circ\at_t(A)\;,$$
exists for sufficiently many $A\in\A$.
This is in general much harder to prove.
But again there is a sufficient condition:
the summability of the Dyson series over all times: for all $A\in\A$,
$$\szi n\iDn{\bnorm{[\a_{-t_n}(V),[\ldots[\a_{-t_1}(V),A]\ldots]]}}<\infty\;.
\eqn$$
\edef\Dyson{\number\eqnno}\noindent
In some cases, including the applications mentioned above,
this Dyson series can be bounded by a sum over rooted trees.
Two examples will be treated in Section \examplesec.
Due to the good combinatorial properties of trees,
we thus obtain all-time integrability of the series in a definite,
non-vanishing regime of perturbations $V$.
This is the subject of Theorem 1.

\sn
The result is an improvement on estimates which were used in
the older cases mentioned above.
It is optimal in the sense that it becomes an equality
if all contributions are non-negative.
We will state it in Section \resultsec,
and give a proof in Section \proofsec.

\goodbreak


\newsection{The main result} \edef\resultsec{\number\secno}\noindent
Let $\seqz m$ and $\seqz\mt$ be
two sequences of nonnegative numbers, and $f,\fet$ two nonnegative
integrable functions on $[0,\infty)$.
Assume that the numbers $\mt_1,\mt_2\ldots$ are not all zero.
We consider the sum of integrals
   $$\eqalign{\Phi(\mt,m,\fet,f)
       &:=\szi n\quad\sCn\mt_{d_c(0)}\left(\pon j \;m_{d_c(j)}\right)\cr
       &\times\iDn{\left(\pon j f_{c_j}(t_j-t_{c_j})\right)},\cr}
   \eqn$$
\edef\Phidef{\number\eqnno}\noindent
where $t_0=0$ and, for $j=0,1,\ldots,n$,
   $$d_c(j):=\#\set{i\in\{1,2,\cdots,n\}}{c_i=j}\eqn$$
\edef\degreedef{\number\eqnno}\noindent
and
   $$f_i:=\cases{\fet&if $i=0$,\cr f&if $i\ge1$.\cr}\eqn$$
\edef\fidef{\number\eqnno}\noindent
Let the generating functions $G,\Gt:[0,\infty)\to[0,\infty]$ be defined by
   $$G(x):=\szi k {{m_k}\over{k!}}x^k\and
   \Gt(x):=\szi k {{\mt_k}\over{k!}}x^k\;,\eqn$$
\edef\eqG{\number\eqnno}
and let $\norm f,\norm\fet$ denote the integrals of $f$ and $\fet$
respectively.

\proclaim Theorem \lemn.
The sum $\Phi(\mt,m,\fet,f)$ in (\Phidef) converges if and only if the equation
   $$G(\|f\|y)=y\eqn$$
allows a solution $y$ for which $\Gt(\norm\fet y)<\infty$.
If $y$ is the least such solution, then
   $$\Phi(\mt,m,\fet,f)=\Gt(\norm\fet y)\;.\eqn$$

\advance\eqnno by -1
\edef\snijpunt{\number\eqnno}
\advance\eqnno by 1
\edef\eqPhi{\number\eqnno}
\edef\estimate{\number\lemno}

\bigskip\bigskip
\appictname{10.5truecm}{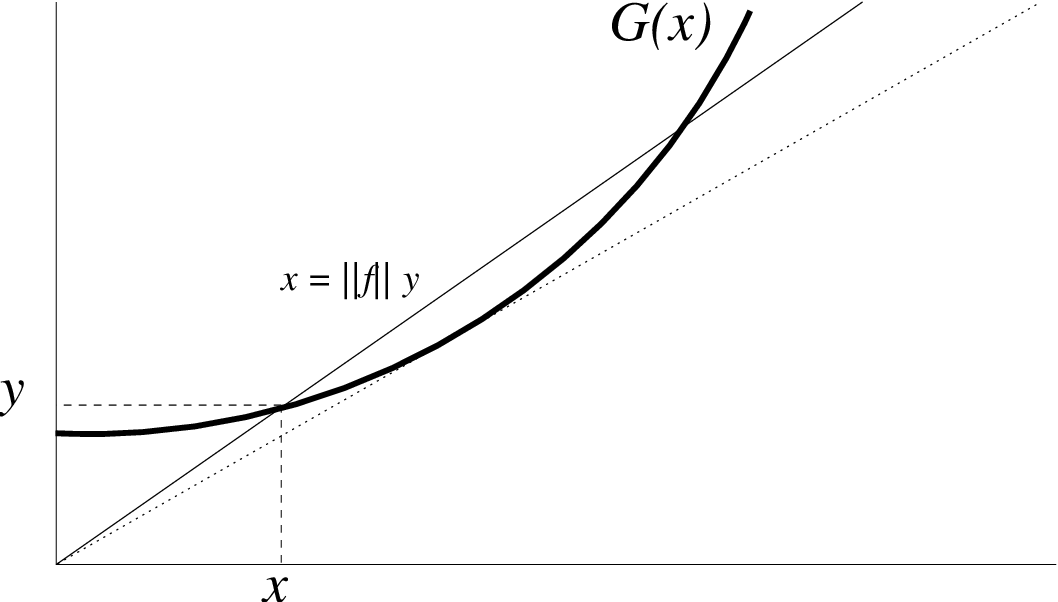}{Fig. 1: the convergence condition.}

\noindent{\bf Discussion}
\medskip\noindent
At first sight,
the chances for convergence of the series (\Phidef) may look slim.
The number of integrals grows as $(n-1)!$,
and the integrals themselves seem to behave roughly like $\norm f^n$.
Hence the older estimates [Maa84] and [Spo97] made a stronger requirement
on $f$ than integrability, namely exponential decrease.
It was clear, however, from work of Botvich, Fayolle and Malyshev [BFM94]
that something better should be possible.
The basic trick of the present paper is to reduce the number of terms
in the sum from factorial to roughly a power law in $n$
by `packing' many terms into a single integral.
(See the `packing lemma', Lemma \packing.)

\noindent
Apart from this integration aspect, there is also the summation
aspect, related to the classical theory of Galton and Watson on family trees,
which makes exact calculation of the sum possible.
We hope to shed some light on these aspects separately
in the following two corollaries to Theorem \estimate.

\proclaim Corollary \lemn.
Let $f:[0,\infty)\to[0,\infty)$ be Lebesgue integrable.
Then for all $n\in\NN$:
  $$\eqalign{\mathrel{\mathop{\int\cdots\int}_{0\le t_1\le\cdots\le t_n}}&
f(t_1)\bigl(f(t_2)+f(t_2-t_1)\bigr)\bigl(f(t_3)+f(t_3-t_1)+f(t_3-t_2)\bigr)\cr
      &\times\cdots\times\bigl(f(t_n)+f(t_n-t_1)+\ldots+f(t_n-t_{n-1})\bigr)
             dt_1 dt_2\ldots dt_n\cr
     &={{(n+1)^{n-1}}\over{n!}}\left(\izi f(t)dt\right)^n\;.\cr}$$

\edef\IntCom{\number\lemno}\noindent
We note that this expression grows roughly like $e^n\norm f^n$,
and is therefore summable over $n$ for $\norm f<{1\over e}$.

\proof
Let us put
$\mt_k=m_k=1$ ($k\in\NN$), $\|f\|=\norm\fet=:u$.
Then $\Gt(x)=G(x)=e^x$, and
Theorem \estimate\ implies that the sum $\Phi(u):=\Phi(\mt,m,\fet,f)$ satisfies
   $$\Phi(u)=e^{u\Phi(u)}\;.$$
\noindent
So, putting $A(u):=u\Phi(u)$ we find
   $$A(u)=ue^{A(u)}.\eqn$$
\noindent
$A$ and $\Phi$ are known as the generating functions of the combinatorial
species `rooted tree' and `forest' in the sense of Joyal [Joy81, BLL98].
$A$ is the inverse of the function  $z\mapsto ze^{-z}$,
which can be found using Lagrange's inversion formula.
The result for $\Phi$ is then
   $$\Phi(u)=\szi n {{(n+1)^{n-1}}\over{n!}}\;u^n\;.\eqn$$
\edef\eqPhit{\number\eqnno}\noindent
In particular, the $n$-th term of the sum in (\Phidef) is
$(n+1)^{n-1}\|f\|^n/n!$.
\qed

\sn
Corollary \IntCom\ shows how Theorem \estimate:
`packs' a series of integrals over ordered $n$-tuples of times
into a weighted sum over rooted trees of the $n$-fold integral over
$[0,\infty)$.

\smallskip\noindent
Let us illustrate the combinatorial mechanisms involved in more detail,
by looking at the case $n=3$ in Corollary \IntCom.
It contains six terms.
For each term we make a tree diagram (Fig. 1) with four vertices,
numbered 0,1,2, and 3, by drawing a line from vertex $i$ to vertex $j$
for each factor $f(t_i-t_j)$,
and a line from $i$ to the ``root'' 0 for each factor $f(t_i)$.

\appictname{12truecm}{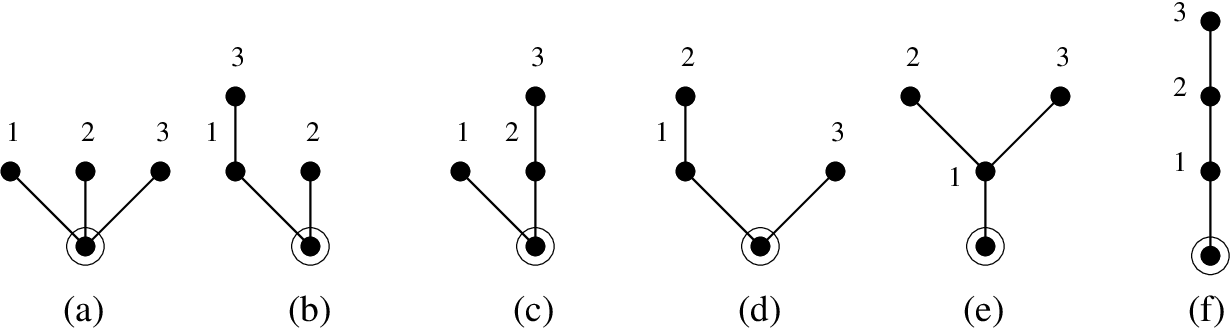}{Fig 1: Rooted trees labeling the integrals.}

\def\If{\left(\int f\right)}\noindent
Now, integral (a) yields ${1\over6}\If^3$,
since it is the integral of a symmetric function over one sixth of
$[0,\infty)^3$:
  $$\iDthree{f(t_1)f(t_2)f(t_3)}
              =\sixth\izi\izi\izi f(t_1)f(t_2)f(t_3)dt_1dt_2dt_3\;.$$
Integral (f) yields $\If^3$ via the change of variables
$u_1:=t_1$, $u_2:=t_2-t_1$, $u_3:=t_3-t_2$.
Integral (e) can be written
  $$\izi f(t_1)\left(\ini{t_1}\ini{t_2}f(t_2-t_1)f(t_3-t_1)dt_3\,dt_2\right)
         dt_1\;,$$
which leads to $\half\If^3$ via a change of variable
$u_2:=t_2-t_1$, $u_3:=t_3-t_1$,
and a symmetry argument similar to that for term (a).
And finally,
integrals (c), (b) and (d) come from one and the same {\it type} of tree.
They can be packed together to yield $\If^3$ as follows:
  $$\izi f(u)\ini u f(s-u)\left(\int_0^u f(t)dt
      +\int_u^s f(t)dt+\ini s f(t)dt\right)ds\,du
      =\left(\izi f(t)dt\right)^3\;.$$

\noindent
So we see here that each {\it type} of tree diagram
yields the amount $\If^3$ divided by the number of its symmetries
(or {\it automorphisms}).
The number of terms reduces from six to four.
Adding the four terms we find, as announced,
   $$\bigl(\sixth+1+\half+1\bigr)\If^3
       ={\textstyle{8\over 3}}\If^3={{(3+1)^{3-1}}\over{3!}}\If^3\;.$$


\sn
The second aspect of Theorem 1
is connected to the Galton-Watson theory of branching processes
(e.g. [Har63]).
In its original form this theory was concerned with the extinction
of family names.

\noindent
Suppose that every male in a family has probability $p_j$
to produce $j$ male offspring in the next generation,
each of which independently produces an identically distributed number of
(male) descendants in the next, etcetera.
If we start with a single individual bearing a certain unique surname,
what is the probability that this surname will eventually die out?
(The Nineteenth Century scholars did not anticipate women's liberation
developments.)

\noindent
Here we don't give the textbook argument, but one which suits our purpose:
Since the alternative to extinction is an {\it infinite} family tree,
the probability of extinction must be the sum over all {\it finite}
family trees of the probabilities of their realization.
It is a small combinatorial puzzle to check that this amounts to 
   $$\sum_{\hbox{types of rooted trees}}{1\over{\#(\hbox{symmetries tree})}}
            \prod_{\hbox{vertices $v$}}d(v)!\,p_{d(v)}\;,$$
where $d(v)$ stands for the number of descendants of the individual
at vertex $v$.

\proclaim Corollary \lemn.
The extinction probability equals the
smallest fixed point of the probability generating function
   $$G(x):=\szi n p_n\,x^n\;.$$

\noindent
This is the familiar result of Galton-Watson theory.

\noindent
In fact, we could even come closer to Theorem \estimate\
by giving the first individual a different offspring distribution from
his decendents.
Indeed, in the quantum mechanical application the root is formed by
the arbitrary observable $A$, and all other nodes represent the perturbation
potential $V$.

\proof
In Theorem \estimate\ put
   $$\mt=m,\qquad\fet=f,\qquad\szi k {{m_k}\over{k!}}=1,
     \and \norm f:=\intRp f(t)dt=1.\eqn$$
\noindent
Then the theorem allows the following interpretation.
In a branching process a single individual
splits at time 0 into $k$ new individuals with probability $m_k/k!$.
These in their turn live for independent random times,
all with probability density $f$,
and produce independent offspring according to the same law $m_k/k!$,
which again live for independent random times, distributed according to $f$,
etcetera.
Then the sum $\Phi(m,m,f,f)$ in (\Phidef)
is the total probability measure carried by all possible finite
family trees, which equals the probability
that the progeny of the original individual will eventually die out.
\qed

\bigskip\noindent{\bf Earlier results}
\medskip\noindent
The case of Corollary \IntCom\ was studied by Botvich,
Fayolle and Malyshev [BFM94] in the context of network theory.
There the $n$-th term was estimated by $(8\|f\|)^n$.
Our result is also, in this special case, slightly better than
the result of [FiL99] which gives for the $n$-th term the estimate
   $$2^n t^n\sqrt e c^{{n\log\log n}\over{\log n}}\;,\quad(c>e)\;.$$
But note that (\eqPhi) and (\eqPhit) are equalities, not inequalities.

\noindent
In [Maa84] and [Spo97] an estimate was given by requiring exponential
decay of $f(t)$ as $t\to\infty$ instead of just integrability.


\newsection{Applications} \edef\examplesec{\number\secno}\noindent
Now, in what kind of situations can we expect the estimate of Theorem
\estimate\ to apply to the Dyson series (\Dyson)?
Let us fix self-adjoint operators $A$ and $V$ in $\A$.

First, it is important that the commutators
   $$[\a_{-t}(V),A] \and [\a_{-t}(V),V]$$ 
tend to zero as $t$ tends to infinity in an integrable way.
This is an instance of a property called
``$L^1$-asymptotic abelianness''.
and expresses the fact that both $A$ and $V$ are local observables,
which eventually get separated by the dynamics $\a_t$.
It leads to the integrability of the functions
$\fet$ and $f$,

Second,
every added commutator $[\a_{-t}(V),\cdot]$,
acting on a repeated commutator
$X=[\a_{-t_n}(V),[\ldots[\a_{-t_1}(V),A]\ldots]]$
already present,
should be bounded by a sum of the form
   $$\sum_{j=1}^n f_j(t-t_j)\norm{X_j'}\;,$$
where $X_j'$ is an expression similar to $X$,
with possibly some minor alteration related to $t_j$,
and where $f_0(t):=\fet(t)$ and $f_j(t-s):=f(t-s)$ for $j\ge1$
are bounds for $\norm{[\a_{-t}(V),A]}$ and $\norm{[\a_{-t}(V),\a_{-s}(V)]}$
respectively.
This formulation is necessarily vague,
every application having its own peculiarities.
Instead of striving for generality,
let us illustrate the transition from the Dyson series in (\Dyson)
to the sum $\Phi(\mt,m,\fet,f)$ in (\Phidef) in two typical cases.

\medskip\noindent{\bf 
Anharmonic oscillator in a bath of oscillators.}\medskip\noindent
Oscillator models are a recurrent theme in the literature
on approach to equilibrium,
from the ancient model of Horace Lamb [Lam00],
via the harmonic chain [Hem79] and the Ford-Kac-Mazur model [FKM65],
to the sophisticated models of Caldeira and Leggett [CaL81],
and Fidaleo and Liverani [FiL99, FiL03].
This class of infinite models is particularly accessible
to analysis due to the fact that their phase spaces are symplectic
vector spaces, and their dynamics are groups of linear symplectic
transformations [LeM84].

For the purpose of quantization the phase space can be
made into a complex Hilbert space $\H$, with the imaginary part of the
inner product as the symplectic form,
and a symplectic group of the form $e^{ith}$ for some
positive definite `one particle' Hamiltonian $h$.
In a mathematically rigorous form,
this approach was pioneered by Segal, Weinless and Kay [Kay79].

The algebra of observables of such an
assembly of coupled harmonic oscillators is described by a
CCR algebra $\A$ over $\H$,
generated by Weyl operators $W(\eta)$,
with $\eta\in\H$, satisfying
   $$W(\eta_1)W(\eta _2)=e^{-{i\over2}\Im\inp{\eta_1}{\eta_2}}W(\eta_1+\eta_2)\;,$$
and with a time evolution $\a_t$ given by
   $$\a_t\bigl(W(\eta)\bigr)=W(e^{ith}\eta)\;.$$
A KMS state on $\A$ at inverse temperatur $\b$ with respect to
this evolution is given by
   $$\oo_\b\bigl(W(\eta)\bigr)=e^{-{1\over4}\inp{\eta}{\coth({1\over2}\b h)\eta}}\;.$$
Via the Gel'fand-Naimark-Segal (GNS) construction this state leads
to a representation of $\A$ on a (much larger) Hilbert space $\H_\b$,
which possesses a distinguished vector $\xi_\b$ reflecting the
KMS state:
   $$\inp{\xi_\b}{A\xi_\b}=\oo_\b(A)\quad\hbox{for}\quad A\in\A\;.$$
The relation to approach to equilibrium is the following:
If $\H$ is infinite-dimensional, which corresponds to infinitely
many oscillators, it may happen that for all $\eta_1,\eta_2\in\H$:
   $$\lim_{t\to\infty}\inp{\eta_1}{e^{ith}\eta_2}=0\;.$$
As a consequence, the dynamics $\a_t$ approaches the equilibrium state
$\oo_\b$ in the sense of (\mixing).
Then the question posed in the introduction arises.
We choose a single oscillator,
say a vector $q\in\H$, whose position operator $Q$ is given by
   $$W(\l q)=e^{i\l Q}\;,$$
and we perturb the dynamics by adding a term $V:=v(Q)$ to the
Hamiltonian on the GNS space.
If $v$ is the Fourier transform of some signed measure $\mu$ on $\RR$,
then
   $$V=v(Q)=\intR e^{i\l Q}\mu(d\l)=\intR W(\l q)\mu(d\l)\;.$$
As discussed in Section \introsec,
we are interested in the uniform convergence of the Dyson series
for this choice of $V$ and $\a_t$.
In fact it suffices to consider $A\in\A$ of the form $A=W(\lambda_0 e^{it_0h}q)$,
for some fixed $t_0\in\RR$ and $\l_0\in\CC$.

To connect up with the sum of integrals $\Phi$ of Section \resultsec,
let $f(t):=|\Im\inp q{e^{ith}q}|$ and $\fet(t):=f(t-t_0)$;
let the measure $\mu_{+}$ be such that $|\mu(S)|\le\mu_{+}(S)$ for all
Borel subsets of $\RR$.
By repeated use of the equality
   $$[W(\eta_1),W(\eta_2)]=-2i\sin(\half\Im\inp{\eta_1}{\eta_2})
      W(\eta_1+\eta_2)\;,$$
we obtain
$$
\eqalign{\bnorm{[&\a_{-t_n}(V),[\cdots[\a_{-t_1}(V),A]\cdots]]}\cr
       &\le\intR\cdots\intR
           \left|\prod_{j=1}^n 2\sin\half
 \left(\sum_{c=0}^{j-1}\Im\inp{\l_je^{it_jh}q}{\l_ce^{it_ch}q}\right)\right|
        \mup(d\l_1)\cdots\mup(d\l_n)\cr
       &\le\sCn\left(\intR\cdots\intR\left(\prod_{j=1}^n|\l_j\l_{c_j}|\right)
                  \mup(d\l_1)\cdots\mup(d\l_n)\right)\cr
 &\qquad\qquad\qquad\qquad\qquad\times
 \left(\prod_{j=1}^n f_{c_j}(t_j-t_{c_j})\right)\;,\cr
       &=\sCns\mt_{d_c(0)}\prod_{j=1}^n m_{d_c(j)}
         \left(\prod_{j=1}^n f_{c_j}(t_j-t_{c_j})\right)\;,\cr}
$$
where $d_c(j)$ and $f_i$ are defined as in (\degreedef) and (\fidef)
in Section \resultsec,
and where $\mt_k:=|\l_0|^k$ and $m_k:=\int_0^\infty|\l|^{k+1}\mup(d\l)$.
Integrating over $t_1,t_2,\ldots,t_n$ and summing over $n$
we obtain the sum $\Phi(\mt,m,\fet,f)$ in (\Phidef).

(It is a pity that, as was shown in [BGM99],
this theory still does not allow the perturbation potential
$v$ to break the convexity of the oscillator potential,
so that metastable oscillator states,
which are an important motivation of this type of oscillator models [CaL81],
still cannot be treated rigorously.)

\medskip\noindent{\bf 
Interacting fermionic open systems.}\medskip\noindent
As our second example,
we shall describe the application of Theorem \estimate\
by Jak\v si\'c, Ogata and Pillet [JOP06]
to prove Green-Kubo formulas and Onsager reciprocity relations
for locally interacting fermionic open systems.
Their model consists of several Fermi gases at different temperatures
and chemical potentials,
which interact by a weak local interaction potential $V$.
Existence and properties of a non-equilibrium stationary state for
the total system are proved by showing uniform convergence of
the Dyson series.
We refer to the paper [JOP06] for further details;
let it suffice here to show how in this case
the Dyson series is estimated by a sum
of integrals of the form $\Phi(\mt,m,\fet,f)$.

Let $\A$ be the C*-algebra generated by creation and annihilation
operators satisfying the {\it canonical
anticommutation relations}
   $$a(\eta_1)a^*(\eta_2)+a^*(\eta_2)a(\eta_1)=\inp{\eta_1}{\eta_2}\cdot\one,$$
valid for all $\eta_1,\eta_2$ in some Hilbert space $\H$.
The unperturbed dynamics $\a_t$ is the automorphism group on $\A$ given by
   $$\a_t\bigl(a(\eta)\bigr)=a\left(e^{ith}\eta\right)\;.$$
The perturbation potential is a (self-adjoint) sum over even monomials
in annihilation and creation operators
   $$V=\l\sum_{k=1}^K a^\#(\th_{k,1})a^\#(\th_{k,2})\cdots a^\#(\th_{k,q_k})$$
for some positive integer $K$, a positive coupling constant $\l$,
and even numbers $q_1,q_2,\ldots,q_K$.
Here $a^\#$ stands for $a$ or $a^*$ and
the vectors $\th_{k,j}$ are chosen from the unit ball of $\H$.
The test observable $A$ is a single monomial of the form
$a^\#(\ph_1)\cdots a^\#(\ph_p)$,
with $p\in\NN$ and $\ph_1,\ldots,\ph_m$ in the unit ball of $\H$.

Now, the commutator of two monomials $A=a_1a_2\cdots a_p$
and $B=b_1b_2\cdots b_q$ of annihilation or creation operators,
at least one of which is of even degree, is calculated as
   $$[B,A]=\sum_{i=1}^p\sum_{j=1}^q(-1)^j(a_ib_j+b_ja_i)
     a_1\cdots a_{i-1}\bigl(b_1\cdots b_{j-1}b_{j+1}\cdots b_q\bigr)
     a_{i+1}\cdots a_p\;.\eqn$$
\edef\monomials{\number\eqnno}\noindent
In particular,
   $$\eqalign{[\a_{-t}(V),A]&=\l\sum_{k=1}^K\sum_{i=1}^p\sum_{j=1}^{q_k}
                   (-1)^j\inp{\th_{kj}}{e^{ith}\ph_i}_*
                         a^\#(\ph_1)\cdots a^\#(\ph_{i-1})\cr
  &\times\left(a^\#\left(e^{-ith}\th_{k1}\right)\cdots
         a^\#\left(e^{-ith}\th_{k,j-1}\right) 
         a^\#\left(e^{-ith}\th_{k,j+1}\right)\cdots
         a^\#\left(e^{-ith}\th_{k,q_k}\right)\right)\cr
  &\times a^\#(\ph_{i+1})\cdots a^\#(\ph_p)\;,\cr}$$
where the starred inner product $\inp{\th_{kj}}{e^{ith}\ph_i}_*$
is to be read as 0 when $a^\#(\ph_i)$ and $a^\#(\th_{kj})$
are both annihilators or both creators,
as the ordinary inner product when the first is an annihilator and the
second a creator,
and as its complex conjugate if it is the other way around.
If we now put $\fet(t):=\l K\max_{i,j,k}|\inp{\th_{kj}}{e^{ith}\ph_i}|$,
and $q:=\max_k q_k$,
then it follows that
   $$\norm{[\a_{-t}(V),A]}\le pq\fet(t)\;.$$
Let us also define an upper bound $f(t)$ for the action of $[\a_{-t},\cdot]$
on $V$
by putting $f(t):=\l K\max_{j,k,j',k'}|\inp{\th_{kj}}{e^{ith}\th_{k'j'}}|$.
Repeated application of (\monomials) then leads to the estimate
   $$\eqalign{&\norm{[\a_{-t_n}(V),[\ldots[\a_{-t_1}(V),A]\ldots]]}\cr
     &\le\l^n\sum_{k_1=1}^K\cdots\sum_{k_n=1}^K\quad\sCn\cr
     &p(p-1)\cdots(p-d_c(0)+1)
      \times\prod_{j=1}^n q_{k_j}\cdots(q_{k_j}-d_c(j))
      \times\prod_{j=1}^nf_{c_j}(t_j-t_{c_j})\;.\cr}$$
where $f_i$ is defined as before in (\fidef).
Summing over $n\in\NN$ and integrating over $t_1,t_2,\ldots,t_n$
with $t_1\le t_2\le\ldots\le t_n$ now yields an estimate of the entire
Dyson series by $\Phi(\mt,m,\fet,f)$, provided we define
   $$\mt_k:=p(p-1)\cdots(p-k+1)\and m_k:=q(q-1)\cdots(q-k)\;.$$
The relevant generating functions for this example are
   $$\Gt(x)=(1+x)^p\and G(x)=q(1+x)^{q-1}\;.$$


\newsection{Proof of the Theorem} \edef\proofsec{\number\secno}\noindent
\smallskip\noindent{\bf Rooted trees}\smallskip\noindent
According to the usual definition a rooted tree is
a finite connected graph without cycles and with one distinguished vertex.
Here we prefer to use the following, equivalent definition.

\noindent
Let $V$ be a finite nonempty set.
A {\it rooted tree} with {\it vertex set} $V$ 
is a function $a:V\to V$ with the property that
there is a point $\odot\in V$ such that for $k\in\NN$ sufficiently large
and all $v\in V$ we have $a^{\circ k}(v)=\odot$.
The point $\odot$ is called  the {\it root} of $a$.
Note that always $a(\odot)=\odot$.
The least value of $k$ for which $a^{\circ k}$
contracts all vertices to the root is the {\it height} of the tree $a$.
By $V^*$ we shall mean $V\setminus\{\odot\}$.
In this paper by a {\it tree} we always mean a rooted tree.

\indpictname{5truecm}{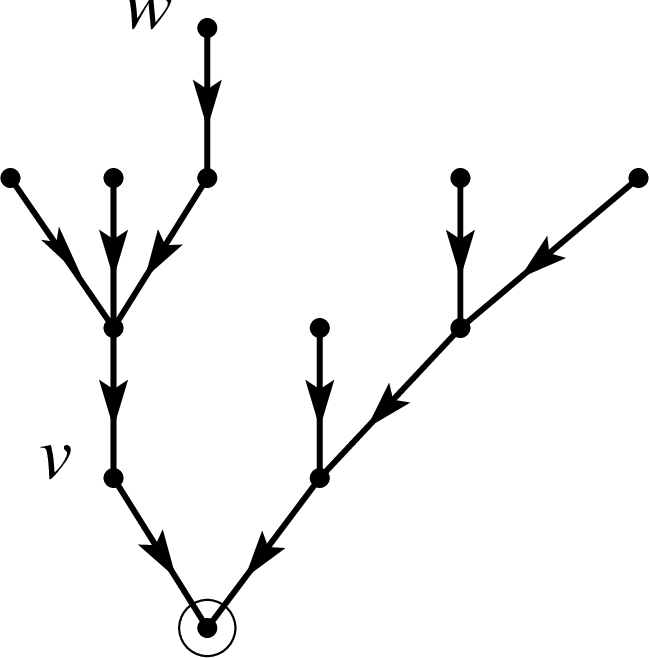}{Fig.2: A rooted tree ($v\prec w$).}{\noindent
Drawing an arrow from $v$ to $a(v)$ for each $v\in V^*$,
we obtain an oriented graph.
The number $n$ of arrows will be called the {\it size} $|a|$
of the tree.
Note that $|a|=\#(V^*)=\#(V)-1$.
The vertex set $V$ is partially ordered by $a$ in a natural way:
we say that $v\prec w$ if $v=a^k(w)$ for some $k\in\NN$.
We think of $a(v)$ as the {\it parent} of $v$
(except, of course, if $v$ is the root).
By $d_a(v)$ we denote the number of points in $V^*$
that are mapped to $v$ by $a$
(the number of {\it children} of $v$ if we regard $a$ as a family tree).}

\pn
Rooted trees $a:V\to V$ and $b:W\to W$ are considered {\it isomorphic}
if there is a bijection $f:V\to W$ such that $b\circ f=f\circ a$.
We denote the collection of all isomorphism classes
or {\it types} of rooted trees by $\AA$.
We write $\AA_n$ ($n\in\NN)$ for the types of trees of size $n$,
and $\AA[h]$ ($h\in\NN$) for the types of trees of height at most $h$.

\pn
We shall denote isomorphism classes of trees by $\a,\b,\ldots$,
and write $|\a|$ for the size of the trees in class $\a$.

\noindent
An {\it automorphism} of a tree $a$ is an isomorphism from $a$ to itself.
We denote the group of all automorphisms of $a$ by $\aut(a)$.
Since all trees of the same type have the same number of automorphisms,
we sometimes write $\#\aut(\a)$ instead of $\#\aut(a)$.
In the same sense we shall speak of $d_\a(\odot)$ for a type $\a$.

\smallskip\noindent{\bf Climbers}\smallskip\noindent
The sum (\Phidef) contains a summation over functions
$c:\{1,\cdots,n\}\to \{0,1,\cdots,n\}$:
$j\mapsto c(j)=c_j$,
which are {\it decreasing} in the sense
that $c(j)<j$ for all $j$.
We shall call such maps {\it climbers} of {\it size} $n$.
Note that, if we add the prescription that $c(0)=0$,
a climber becomes the same as a rooted tree with
vertex set $\{0,1,\cdots,n\}$ satisfying the extra requirement
   $$i\prec j\Implies i\le j\;.$$
We denote the set of all climbers of size $n$ by $\C_n$.

\noindent
We are now going to replace the sum over $\C_n$ occurring in (\Phidef)
by a sum over indexed rooted trees.

\noindent
By an {\it indexation} of a rooted tree $a$ on $V$ of size $n$ we mean
an order-preserving bijection $\iota:V\to\{0,1,2,\ldots ,n\}$.
The set of all indexations of $a$ will be denoted by $I(a)$.
Note that for $\iota\in I(a)$ we always have $\iota(\odot )=0$.

\smallskip\noindent
By indexation a rooted tree becomes a climber:
$c=\iota\circ a\circ\iota^{-1}$. 
There may be more than one indexation leading to the same climber,
as is illustrated in Fig. 3.

\noindent
If $a$ is itself a climber, then $I(a)$ is the set of all isomorphisms
of $a$ with isomorphic climbers $c$.

\bigskip
\appictname{12truecm}{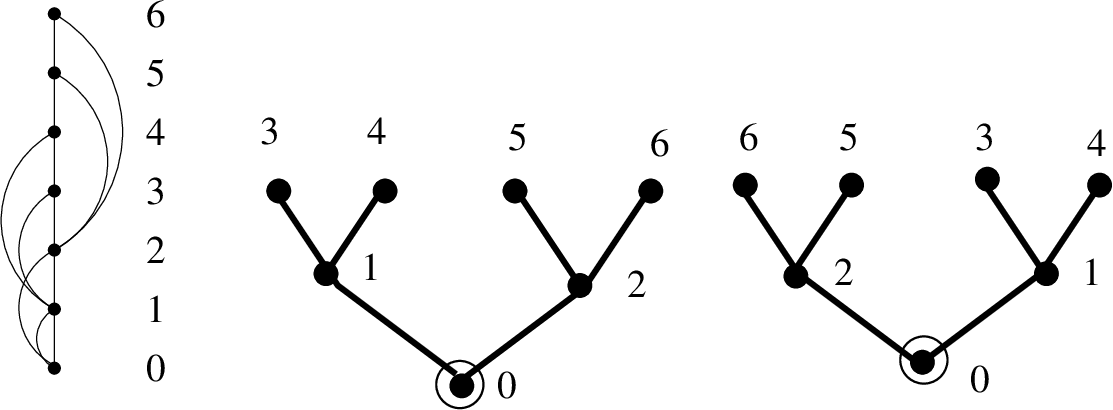}{Fig. 3. Two indexations of a rooted tree
leading to the same climber.}
\bigskip

\proclaim Lemma \lemn.
Let $F:\C_n\to\RR$.
Then
   $$\sum_{c\in\C_n}F(c)
    =\sum_{\a\in\AA_n}{1\over{\#\aut(\a)}}
     \sum_{\iota\in I(a)}F(\iota\circ a\circ\iota^{-1})\;,$$
\nobreak
where $a$ is any tree of type $\a$.

\goodbreak
\edef\climbers{\number\lemno}\noindent
\proof
The set of climbers $\C_n$ decomposes into isomorphism classes
$\a\cap\C_n$ with $\a\in\AA_n$.
The sum over one such class can be performed by summing over the orbit
of a single element $c\in\a\cap\C_n$ under the group of isomorphisms
among climbers ---
which is equal to the set $I(c)$ of indexations of $c$ as noted above ---
and then dividing by the number of automorphisms of $c$.
Replacing $c\in\a\cap\C_n$ by an arbitrary tree $a\in\a$ we obtain
the statement.
\qed

\mn
Now let
   $$\D_n:=\set{t=(\tuple t_n)\in\RR^n}{0<t_1<\cdots<t_n}\;.$$
The following lemma allows us to replace the summation over $I(a)$ together
with the integration over $\D_n$ by a single integral over $[0,\infty)^n$.

\proclaim {Lemma \lemn.} (Packing Lemma).
For any rooted tree $a$ with index set $V$ the map
   $$\th_a:\quad I(a)\times\D_n\tto[0,\infty)^{V^*}:
           \quad (\iota,t)\mapsto r,
           \quad\hbox{where}\quad r_v:=t_{\iota(v)}-t_{\iota(a(v))},$$
(with $t_0=0$)
is bijective up to a subset of $[0,\infty)^{V^*}$ of measure zero,
and has Jacobian 1 on each component $\{\iota\}\times\D_n$.

\edef\packing{\number\lemno}\noindent
\proof
Choose a point $r\in[0,\infty)^{V^*}$ and put $r_\odot:=0$.
Allocate a `branching time' $s_v$ to each vertex $v\in V$ by putting
   $$s_v:=\sum_{w\prec v}r_w\;.$$
If some of these values $s_v$ coincide,
which happens only for a set of points $r$ of measure 0,
then $r$ is not in the range of $\th_a$.
If they are all different,
they determine by their order a unique indexation $\iota$ of $V$:
   $$s_v\le s_w \Equivalent \iota(v)\le\iota(w).$$
Putting $t_{\iota(v)}:=s_v$ we obtain $t\in\D_n$ with the property that
   $$\th_a(\iota,t)_v:=t_{\iota(v)}-t_{\iota(a(v))}
                      =s_v-s_{a(v)}
                      =\sum_{w\prec v}r_w-\sum_{w\prec a(v)}r_w=r_v.$$
So $r$ lies in the range of $\th_a$.
Conversely,
if $r=\th_a(\iota,t)$,
we must have
   $$t_{\iota(v)}=r_v+t_{\iota(a(v))}=r_v+r_{a(v)}+t_{\iota(a\circ a(v))}
     =\cdots=\sum_{w\prec v}r_w=s_v.$$
And since $0<t_1<t_2<\cdots<t_n$,
the indexation $\iota$ is uniquely determined
by the order of the `branching times' $s_v$, hence by $r$.
So $\th_a$ is injective as well as almost surjective.

\noindent
Finally, the map $t\mapsto\th_a(\iota,t)$, for $\iota\in I(a)$ fixed,
can be written as a $V^{*}\times n$-matrix $(M_{v,k})$, which has
$1$'s at the positions $(v,k)$, where $\iota(v)=k$,
$(-1)$'s at the positions $(v,k)$ with $\iota(a(v))=k$
and $0$'s everywhere else.
We may put $M$ in standard form by ordering the points
in $V^{*}$ according to the indexation $\iota$,
thus putting all the $1$'s on the diagonal (cf. Fig. 4).

\indpict{4truecm}{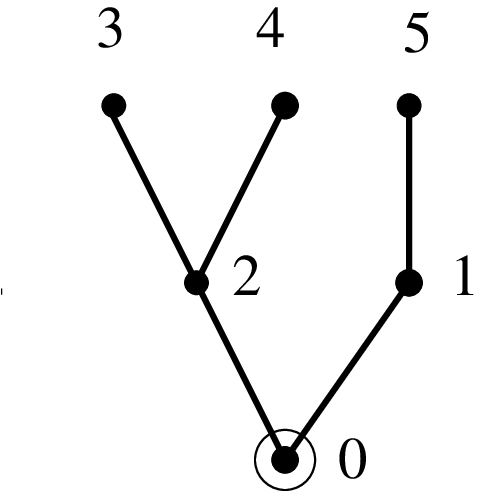}{$$\matrix{1\cr2\cr3\cr4\cr5\cr}
     \left(\matrix{1 & 0 & 0 & 0 & 0 \cr
                   0 & 1 & 0 & 0 & 0 \cr
                   0 &-1 & 1 & 0 & 0 \cr
                   0 &-1 & 0 & 1 & 0 \cr
                  -1 & 0 & 0 & 0 & 1 \cr}\right)$$}

\bigskip\centerline{Figure 4.
A indexed rooted tree $(a,\iota)$ with its matrix $(M_{v,k})$.}

\mn
Then since $\iota(a(v))<\iota(v)$,
all the $(-1)$'s end up below the diagonal.
So $\det(M)$ equals $1$ in this standard form,
and $\pm 1$ in any other ordering of $V^{*}$.
The Jacobian $|\det(M)|$ of the piecewise linear map
$\th_a$ equals $1$ almost everywhere.

\qed

\proclaim Lemma \lemn.
The sum of integrals $\Phi(\mt,m,\fet,f)$ in (\Phidef) can be written as
   $$\Phi(\mt,m,\fet,f)=\sum_{\a\in\AA}
                {{\mt_{d_\a(\odot)}\norm\fet^{d_\a(\odot)}}
                \over{\#\aut(\a)}}
               \left(\prod_{v\in V^*}m_{d_a(v)}
               {\norm f}^{d_\a(v)}\right),$$
where the tree $a$ with vertex set $V$ is any representative of the class $\a$.

\edef\reducPhi{\number\lemno}
\proof
In (\Phidef) $\Phi(\mt,m,\fet,f)$ is written as a sum over $n$
of sums of integrals over
$\C_n\times\D_n$.
First we apply Lemma \climbers\ to
replace the sum over $\C_n$ by a sum over indexed rooted trees,
to obtain
$$
\eqalign{\Phi(\mt,m,\fet,f)&
    =\szi n \sum_{\a\in\AA_n}{{\mt_{d_\a(\odot)}}\over{|\aut(\a)|}}
               \left(\prod_{v\in V^*}m_{d_a(v)}\right)\cr
        &\times\sum_{\iota\in I(a)}\int_{\D_n}\left(\prod_{v\in V^*}
               f_{\iota(a(v))}(t_{\iota(v)}-t_{\iota(a(v))})\right)dt,\cr}
$$
Where again $a:V\to V$ is any tree of type $\a$.
Then we apply the `packing lemma' (Lemma \packing) to replace the sum
over $\iota$ and the integration over $t$ by an integration over $r$:
   $$\Phi(\mt,m,\fet,f)=\sum_{\a\in\AA}{{\mt_{d_\a(\odot)}}\over{|\aut(\a)|}}
           \left(\prod_{v\in V^*}m_{d_a(v)}\right)
           \int_{[0,\infty)^{V^*}}\left(\prod_{v\in V^*}f_{a(v)}(r_v)\right)
           \;dr,$$
where $a:V\to V$ is any tree of type $\a$, and for $x\in V$:
$f_x:=\fet$ if $x$ is the root, otherwise $f_x:=f$.
The integral over $r$ is now easily obtained, and the Lemma is proved.
\qed

\noindent
Now for $\mt,m,\fet$ and $f$ fixed,
let the {\it weight} $\wt(a)$ of a rooted tree $a$,
be as it occurs in Lemma \reducPhi:
   $$\wt(a):=\mt_{d_a(\odot)}\norm\fet^{d_a(\odot)}
             \prod_{v\in V^*}m_{d_a(v)}\norm f^{d_a(v)}.$$
By $w(a)$ we shall denote the same weight, but with $\fet=f$ and $\mt=m$,
which simplifies to
   $$w(a)=\prod_{v\in V}m_{d_a(v)}\norm f^{d_a(v)}.$$
Both $w(a)$ and $\wt(a)$ depend only on the type of $a$,
hence we may write $w(\a)$ and $\wt(\a)$ respectively.

\noindent
 Let $\AA[h]$ with $h\in\NN$ denote the set of all rooted trees
of height $\le h$, and define
   $$\Phi_h(\mt,m,\fet,f):=\sum_{\a\in\AA[h]}{{\wt(\a)}\over{|\aut(\a)|}}\;.$$

\proclaim Lemma \lemn.
For all pairs of sequences $\mt,m$ of nonnegative numbers,
all pairs of integrable functions $f,\fet:[0,\infty)\to[0,\infty)$
and all $h\in\NN$,
   $$\Phi_{h+1}(\mt,m,\fet,f)=\Gt\bigl(\norm\fet\Phi_h(m,m,f,f)\bigr),$$
where $G$ and $\Gt$ are the generating functions given in (\number\eqG).

\edef\recurPhi{\number\lemno}
\proof
If $a$ is a rooted tree of height at most $h+1$
and root degree $d_a(\odot)=p$,
and we cut off its root, then we are left with a set of
$p$ rooted trees of height at most $h$.
On the level of isomorphism classes this leads to a one-to-one correspondence
between $\AA[h+1]$ and $p$-multisets from $\AA[h]$, i.e.
functions $\mu:\AA[h]\to\NN$ which satisfy
   $$\sum_{\b\in\AA[h]}\mu(\b)=p\;.$$
Under this correspondence we have
   $$\wt(\a)=\mt_p\norm\fet^p\cdot\prod_{\b\in\AA[h]}w(\b)^{\mu(\b)}\;,$$
and
   $$\#\aut(\a)=\prod_{\b\in\AA[h]}\mu(\b)!\bigl(\#\aut(\b)\bigr)^{\mu(\b)},$$
since any automorphism of a tree of type $\a$ amounts to a permutation
of isomorphic subtrees, plus the application of an automorphism to each of them.
We calculate, starting from Lemma \reducPhi,
and applying the multinomial formula to $\mu$,
$$\eqalign{\Phi_{h+1}(\mt,m,\fet,f)&
           =\sum_{\a\in\AA[h+1]}{{\wt(\a)}\over{\#\aut(\a)}}\cr
 &=\szi p \mt_p\norm\fet^p\sum_{\scriptstyle\mu:\AA[h]\to\NN
                                       \atop\scriptstyle\sum\mu=p}\quad
   \prod_{\b\in\AA[h]}
   {{w(\b)^{\mu(\b)}}\over{\mu(\b)!(\#\aut(\b))^{\mu(\b)}}}\cr
 &=\szi p\mt_p\norm\fet^p{1\over{p!}}
   \left(\sum_{\b\in\AA[h]}{{w(\b)}\over{\#\aut(\b)}}\right)^p\cr
 &=\Gt\bigl(\norm\fet\Phi_h(m,m,f,f)\bigr)\;.\cr}$$
\nobreak\qed
\goodbreak

\proofof Theorem \estimate.
Let $G_f$ and $\Gt_\fest$ denote the maps
    $$y\mapsto G(\norm f y)\and y\mapsto \Gt(\norm\fet y)\;.$$
By Lemma \recurPhi, and since $\Phi_0(m,m,f,f)=m_0=G_f(0)$,
we have for all $h\in\NN$:
    $$\Phi_{h}(\mt,m,\fet,f)=\Gt_\fest\circ G_f^{\circ h}(0).\eqn$$
\edef\mainseq{\number\eqnno}\noindent
We must study convergence of this expression as $h\to\infty$.
First consider $y_h:=G_f^{\circ h}(0)$.
Suppose that (\snijpunt) has a solution,
i.e. $G_f$ has a fixed point $u\ge0$.
Then, since $G_f$ is non-decreasing, and since $0\le u$,
we have for all $h\ge1$:
   $$y_h=G_f^{\circ h}(0)\le G_f^{\circ h}(u)=u.$$
Being bounded above, the sequence converges to a limit $y\le u$.
As $y$ must be a fixed point itself, it is the least such point.

\noindent
On the other hand, if (\snijpunt) has no solution,
the sequence $y_1,y_2,y_3,\cdots$ can have no finite limit,
so it must tend to infinity.

\noindent
Finally, consider the sequence
$\bigl(\Gt_\fest(y_h)\bigr)_{h\in\NN}$ in (\mainseq).
By assumption, $\mt_1,\mt_2,\ldots$ are not all zero, hence
$\Gt_\fest$ is strictly increasing and convex.
This implies that the limit
   $$\Phi(\mt,m,\fet,f)=\li h\Phi_h(\mt,m,\fet,f)=\li h\Gt_\fest(y_h)$$
exists iff the increasing sequence $y_h$ converges
to a point $y$ in the region of convergence of $\Gt_\fest$.
As argued above, $y$ is the least fixed point of $G_f$ if such exists.
We conclude that the sequence (\mainseq) converges if and only if the
least fixed point $y$ of $G_f$ lies in the domain of $\Gt_\fest$;
in that case, 
   $$\Phi(\mt,m,\fet,f)=\Gt_\fest(y)\;.$$
\qed

\acknowledgement
We thank the anonymous referee for corrections and valuable suggestions
for improvement.

\references
\item{[BFM94]}
Botvich,~D., Fayolle,~G., Malyshev,~V.:
``Loss Networks in Thermodinamic Limit'',
In: {\sl 11th Conf. Analysis and Optimisation of Systems:
discrete event systems},
Sopia-Antipolis France 1994,
Springer Verlag Berlin
Lect. Notes Control Inf. Sci. {\bf 199}, (1994), 465--489. 

\item{[BGM99]}
Botvich,~D., Gu\c t\u a,~M., Maassen,~H.:
Stability of Bose dynamical systems and branching theory'',
Preprint (mp-arc 99-130).

\item{[BoM83]}
Botvich,~D., Malyshev,~V.A.:
``Unitary equivalence of temperature dynamics for ideal and locally perturbed
Fermi gas'',
Commun. Math. Phys. {\bf 91} (1983), 301-312.

\item{[BLL98]}
Bergeron,~F., Labelle,~G., Leroux,~P.:
``Combinatorial species and tree-like structures'',
Cambridge University Press 1998.

\item{[CaL81]}
Caldeira~A.C., Leggett~A.J.:
``Influence of dissipation on quantum tunneling in microscopic systems'',
Phys. Rev. Lett. {\bf46} (1981), 211--214.

\item{[FiL99]}
Fidaleo,~F., Liverani,~C.:
``Ergodic Properties for a Quantum Non Linear Dynamics'',
J.~Stat.~Phys. {\bf 97} (1999), 957--1009.

\item{[FiL03]}
Fidaleo,~F., Liverani,~C.:
``Ergodic properties of a model related to disordered quantum anharmonic
crystals'',
Commun.~Math.~Phys. 235 (2003), 169--189. 

\item{[FKM65]}
Ford,~G., Kac,~M., Mazur,~P.:
``Statistical Mechanics of Assemblies of Coupled Oscillators'',
J.~Math.~Phys. {\bf 6} (1965), 504--515.

\item{[FMU04]}
Fr\"ohlich,~J., Merkli,~M., Ueltchi,~D.:
``Dissipative transport: thermal contacts and tunneling junctions'',
Ann. Henri Poincar\'e {\bf 4} (2004), 897--945.

\item{[Har63]}
Harris, T.E.:
``The theory of branching processes'',
Springer, Berlin 1963.

\item{[Hem79]}
J.L. van Hemmen,
``Dynamics and ergodicity of the infinite harmonic crystal,
a review of some salient features'',
Springer Lecture Notes in Physics, Vol. 93 (1979)

\item{[JOP06]}
Jak\u si\'c,~V., Ogata,~Y., Pillet,~C.-A.:
``The Green-Kubo formula for locally interacting fermionic open systems'',
{\it to appear in} Ann. Henri Poincar\'e.

\item{[Joy81]}
Joyal,~A:
``Une Th\'eorie Combinatoire des Series Formelles'',
Adv. Math. {\bf 42}, (1981), 1--82.

\item{[Kay79]}
Kay, B.S.:
``A uniqueness result in the Segal-Weinless approach to linear Bose
fields'',
J. Math. Phys. {\bf 20} (1979), 1712--1713.

\item{[Lam00]}
H.~Lamb,
``On a peculiarity of the wave-system due to the free vibrations
of a nucleus in an extended medium'',
Proc. Lond. math. Soc. {\bf 2} (1900) 88.

\item{[LeM84]}
Lewis,~J.T., Maassen,~H.:
``Hamiltonian Models of Classical and Quantum Stochastic Processes'',
In: {\sl Quantum Probability and Applications to the Quantum Theory of
Irreversible Processes}, Lect. Notes Math. {\bf 1055}, Springer 1984.

\item{[Maa82]}
Maassen,~H.:
``On the invertibility of M\o ller morphisms'',
J.~Math.~Phys. {\bf 23} (1982), 1848--1851.

\item{[Maa84]}
Maassen,~H.:
``Return to Equilibrium by a Solution of a Quantum Langevin Equation'',
J.~Stat.~Phys. {\bf 34}, (1984), 239--262.

\item{[Rob73]}
Robinson,~D.W.:
``Return to equilibrium'',
Commun.~Math.~Phys.{\bf 31} (1973), 171--189.

\item{[Spo97]}
Spohn,~H.:
``Asymptotic Completeness for Rayleigh Scattering'',
J.~Math.~Phys. {\bf 38}, (1997), 2281--2296.

\bye